\documentclass[12pt]{iopart}

\usepackage{graphics}
\usepackage{epsfig}
\usepackage[T1]{fontenc}
\usepackage{float}
\usepackage{lineno}
\usepackage{cite}

\begin{document}
\title[Author guidelines for IOP Publishing journals in  \LaTeXe]{Imagining density distribution of molecular orbitals in IR+XUV co-rotating circular laser fields by frequency-domain theory}

\author{Yu-Hong Li$^{1,8}$, Facheng Jin$^{2,3,*}$,Yujun Yang$^{4}$, Fei Li$^{3}$, Ying-Chun Guo$^{5}$,  Zhi-Yi Wei$^{1,8,9}$, Jing Chen$^{6}$, Xiaojun Liu$^{7}$ and Bingbing Wang$^{1,8,*}$}

\address{$^1$Laboratory of Optical Physics, Beijing National Laboratory for Condensed Matter Physics, Institute of Physics, Chinese Academy of Sciences, Beijing 100190, China}
\address{$^2$Faculty of Science, Xi’an Aeronautical University, Xi’an 710077, China}
\address{$^3$Research Center for Advanced Optics and Photoelectronic, Department of Physics, College of Science, Shantou University, Shantou, Guangdong 515063, China}
\address{$^4$Insititute of Atomic and Molecular Physics, Jilin University, Changchun 130012, China}
\address{$^5$Department of Physics, School of Physics and Electronics, East China Normal University, Shanghai 200062, China}
\address{$^6$Department of Modern Physics, and Hefei National Research Center for Physical Sciences at the Microscale and School of Physical Sciences, University of Science and Technology of China, Hefei 230026, China}
\address{$^7$State Key Laboratory of Magnetic Resonance and Atomic and Molecular Physics, Wuhan Institute of Physics and Mathematics, Chinese Academy of Sciences, Wuhan 430071, China}
\address{$^8$University of Chinese Academy of Sciences, Beijing 100049, China}
\address{$^9$Songshan Lake Materials Laboratory, Dongguan 523808, China}

\ead{fchjin@163.com}
\ead{wbb@aphy.iphy.ac.cn}
\vspace{10pt}
\begin{indented}
\item[]October 2023
\end{indented}

\begin{abstract}
We have investigated the angle-resolved ATI spectrum of oriented molecules in the IR+XUV co-rotating circular laser fields. According to the different roles of IR and XUV laser in the ionization process, we purposefully adjust the photon energy of XUV and the intensity of IR laser to make the ionization spectrum of the molecule distributed in a suitable momentum region. Moreover, under the same laser conditions, the background fringes in the ionization spectrum of the molecule can be removed by using the ionization spectrum of the atom with the same ionization energy as the molecule, so that the molecular orbital density distribution in the suitable momentum region can be obtained. That is, for any unknown molecule, as long as the ionization energy of the molecule can be measured, the density distribution of the molecular orbital can be imaged in a definite momentum region by adjusting the laser field conditions, which may shed light on the experimental detection of molecular orbitals.
\end{abstract}

%
\vspace{2pc}

\section{Introduction}

When an atom or molecule is exposed to a short intense laser pulse, the electron is either freed directly or driven back to the parent ion which may lead to many recollision physical phenomena. The directly ionized or re-scattered electrons carry the information of the initial state of the target atoms or molecules, hence by analyzing the photoelectron signals \cite{meckel2008laser,blaga2012imaging,pullen2015imaging,peters2011ultrafast} or harmonic emission \cite{itatani2004tomographic,chen2013asymmetric, zhong2022high}, the information of atomic or molecular structures and their ultrafast dynamics can be extracted. It has been demonstrated that laser-induced electron tunneling spectroscopy (LETS) can be used to image molecular orbitals, where the density distribution of the molecular orbital can be obtained by scanning the ionization rate of molecules in different directions \cite{meckel2008laser, tong2002theory,lin2006effects}. On the other hand, laser-induced electron diffraction (LIED) technology can realize the self-imaging of molecular orbital during strong-field-induced recollision, which can image sub-ångström structural changes in molecules with femtosecond time resolution \cite{blaga2012imaging,pullen2015imaging,peters2011ultrafast}. Recently, the initial probability distribution 
~\cite{he2015ionization} and the structural information ~\cite{shi2019angle,zhou2020interference} of molecules were imaged from the angular distribution of the directly ionized electron, where the extreme ultraviolet (XUV) laser fields were applied. 

With the rapid development of free-electron laser (FEL) technology ~\cite{shintake2008compact,emma2010first,schoenlein2019recent} and the application of high-order harmonic generation (HHG) ~\cite{heslar2017probing,berman2018coherent}, the XUV and infrared (IR) two-color laser fields have been an important tool to investigate the ionization of electrons with their flexible operability and rich diversity ~\cite{kim2005highly,zhang2013terracelike,hoang2017retrieval,jin2018angle,li2023internal}. Compared with the two-color linearly polarized laser fields, the two-color circularly polarized laser fields provide an additional parameter: relative helicity.  Especially, the so-called bicircular laser fields are composed of two coplanar circularly polarized laser fields with different frequencies and the same or opposite helicities, where these fields have raised increasing interest since it was found experimentally that HHG and attosecond pulses generated in the counter-rotating two-color (CRTC) fields are circularly polarized \cite{eichmann1995polarization,long1995model,fleischer2014spin,qiao2023modulation}. 
In recent years, the CRTC fields
and corotating two-color (CoRTC) fields have been used to control nonsequential double ionization (NSDI) \cite{mancuso2016controlling,eckart2016nonsequential}, investigate nonadiabatic offsets of the initial electron momentum distribution in strong-field tunnel ionization \cite{eckart2018direct}, and observe experimentally subcycle interference structures in the photoelectron momentum distributions (PEMDs) \cite{eckart2018subcycle,eckart2020sideband}, etc. Additionally, it has been shown that PEMDs generated by bicircular laser fields carry the information about the molecular orientation and structure \cite{yuan2017molecular,wang2023comparative,habibovic2018molecules}.

In this work, we investigate the direct above-threshold ionization (ATI) process of the molecule in two co-rotating IR+XUV circularly polarized laser fields by using the frequency-domain theory based on the nonperturbative quantum electrodynamics \cite{guo1992stationary,guo1988quantum}.
The frequency-domain theory has been developed and successfully applied to deal with many strong-field processes since it was first published by Guo et al. in 1988 \cite{guo1988quantum}, where the physics processes include ATI \cite{guo1989scattering,PhysRevA.44.3169,zhang2013terracelike,fu2001interrelation, jin2021influence,zhou2020interference}, HHG  \cite{fu2001interrelation, gao2000nonperturbative}, high-order above-threshold ionization (HATI) \cite{wang2007frequency,guo2009imaging,wang2010charge,jin2018angle} and NSDI \cite{wang2012frequency,jin2016nonsequential,jin2016angle}, etc. Especially, Fu et al. demonstrated that the time-dependent formulas of HHG  can be derived directly from the corresponding formulas of the frequency-domain theory, they hence established the relationship between the frequency- and the time-domain theory \cite{fu2001interrelation}. Moreover, compared with the time-domain theory, the frequency-domain theory provides us with a different quantum-transition viewpoint on the interaction between matter and intense laser fields which has the following advantages: (i) for the ionization processes in  IR+XUV two-color laser fields, this approach can separate various ionization channels to demonstrate the roles that the XUV and IR fields play in the ionization processes; (ii) all of the strong-field processes can be investigated under a unified theoretical frame, which is in favor of analyzing the relationship among all these processes; (iii) another advantage of this approach is avoiding the time-evolution calculation, which can greatly save computation time.

In this paper, we demonstrate that the ATI momentum spectrum of molecules in different energy ranges can be obtained by changing the laser conditions. Therefore, the density distribution of the molecular orbital in the momentum spectrum can be imaged by the angle-resolved ATI spectrum in an appropriate energy range. Atomic units are used throughout unless stated otherwise.

\section{Theoretical method}
The frequency-domain theory for an atom exposed to a two-color laser field has been presented in details \cite{zhang2013terracelike,jin2021influence}. Here this method is briefly summarized and applied to molecule systems. Based on the frequency-domain theory, the Hamiltonian of the molecule-laser system can be written as

\begin{equation}\label{eq1}
H=H_0+U+V \ ,
\end{equation}
where ${H_0} = \frac{{{{\left( { - i\nabla } \right)}^2}}}{{2{m_e}}} + {\omega _1}{N_{{a_1}}} + {\omega _2}{N_{{a_2}}}$represents the energy operator for a free electron-photon system, ${\omega _s}{N_{as}} = {\omega _s}\frac{1}{2}({a_s}^ + {a_s} + {a_s}{a_s}^ + )$  is the energy operator of the laser field with frequency ${\omega _s}$, with ${a_s}$ and ${a_s}^ + $ being the annihilation and creation operators of the photon mode for \emph{s}=1, 2. \emph{U} represents the molecular binding potential, and \emph{V} is the electron-photon interaction potential expressed by
\begin{equation}\label{eq2}
\fl V = \! \! - \frac{e}{{2{m_e}}}\left\{ {\left( { - i\nabla } \right) \cdot [{{\bf{A}}_1}({\bf{r}}) \! + \! {{\bf{A}}_2}({\bf{r}})] \! + \! [{{\bf{A}}_1}({\bf{r}}) \! + \! {{\bf{A}}_2}({\bf{r}})] \cdot \left( { - i\nabla } \right)} \right\} + \frac{{{e^2}}}{{2{m_e}}}{\left[ {{{\bf{A}}_1}({\bf{r}}) \! + \! {{\bf{A}}_2}({\bf{r}})} \right]^2},
\end{equation}
where ${{\bf{A}}_s}{\bf{(r)}} = {g_s}({\hat \varepsilon _s}{e^{i{\bf{k}} \cdot {\bf{r}}}}{a_s} + c.c.)$ is the vector potential with ${g_s} = {(2{\omega _s}{V_s })^{ - 1/2}}{\kern 1pt} {\kern 1pt} (s = 1,2)$  and ${V_s }$  is the normalized volume of the laser field. The polarization vector of the laser field is defined as $\hat \varepsilon_s  = {\hat \varepsilon _x}\cos (\xi_s /2) + i{\hat \varepsilon _z}\sin (\xi_s /2)$ , where $\xi_s$ determines the polarization degree of the laser field with $s = 1,2$, such as $\xi_s  = \pi /2$ corresponds to circular polarization and $\xi_s  = 0$  corresponds to linear polarization. 

The frequency-domain theory based on the Hamiltonian Eq.~(\ref{eq1}) enables us to treat an atom-laser system as an isolated system that consists of an atom and photons. Since the total energy of the system is conserved throughout the interaction process, the formal scattering theory \cite{gell1953formal} can be applied. The \textit{S}-matrix element between the states of the system is \cite{gell1953formal}
\begin{equation}\label{eq3}
{S_{fi}} = \left\langle {\psi_f^-} \right|\left. {\psi_{_i}^+} \right\rangle,
\end{equation}
where
\begin{equation}\label{eq4}
\psi_{_i}^ +  = {\psi_i} + \frac{1}{{{E_i} - H + i\varepsilon }}V{\psi_i}\ ,
\end{equation}
and
\begin{equation}\label{eq5}
\psi_{_f}^- = {\psi_f} + \frac{1}{{{E_f} - H - i\varepsilon }}U{\psi_f}\ .
\end{equation}
Physically, $\psi_{_i}^+$  is the state at \emph{t}=0  which develops from a precollsion state ${\psi_i}$  in the remote past, while $\psi_{_f}^-$  is the state at \emph{t}=0  which evolves to a postcollision state ${\psi_f}$ in the remote future. The \emph{S}-matrix element can be expressed as
\begin{equation}\label{eq6}
{S_{fi}} = {\delta_{fi}} - 2\pi i\delta ({E_f} - {E_i}){T_{fi}}\ .
\end{equation}
Here
\begin{equation}\label{eq7}
{T_{fi}} = \left\langle {{\psi_f}} \right|V\left| {\psi_i^+} \right\rangle  = \left\langle {{\psi_f}} \right|V\left| {{\psi_i}} \right\rangle  + \left\langle {{\psi _f}} \right|U\frac{1}{{{E_i} - H + i\varepsilon }}V\left| {{\psi_i}} \right\rangle\ ,
\end{equation}
is the transition matrix element. Here the expansions of \emph{T}-matrix include two terms, where the first and second terms indicate the ATI and HATI processes respectively. The initial quantum state is expressed as $\left| {{\psi _i}} \right\rangle = {\Phi _i}({\bf{r}}) \otimes \left| {{l_1}} \right\rangle  \otimes \left| {{l_2}} \right\rangle$, which is the eigenstate of the Hamiltonian $H{}_0 + U$ with the associated energy ${E \rm_{i}} =  - {E \rm_{B}} + ({l_1} + \frac{1}{2}){\omega _1} + ({l_2} + \frac{1}{2}){\omega _2}$, ${\Phi _i}({\bf{r}})$ is the ground state wave function of the molecule with the binding energy ${E \rm_{B}}$, $\left| {{l_1}} \right\rangle$ and $\left| {{l_2}} \right\rangle$ are the Fock states of the two laser modes. The final state of the system can be denoted by $\left| {{\psi_f}} \right\rangle  = \left| {{\psi_{{{\bf{P}}_f}{{{m}}_1}{m_2}}}} \right\rangle$, which is the quantized-field Volkov state in two-color laser fields that can be expressed as ~\cite{guo1992stationary}
\begin{equation}\label{eq8}
\begin{array}{l}
\left| {{\psi _{{\bf{P}}_{f}}}_{{m_1}{m_2}}} \right\rangle  = {V_e}^{ - 1/2}\exp [i({\bf{P}}_{f}{\rm{ + }}{u_{{P_1}}}{{\bf{k}}_1} + {u_{{P_2}}}{{\bf{k}}_2}) \cdot {\bf{r}}]\sum\limits_{\scriptstyle{q_1} =  - {m_1}\hfill\atop
\scriptstyle{q_2} =  - {m_2}\hfill}^\infty  {{\Im _{{q_1},{q_2}}}{{(\zeta )}^*}} \\
\; \times \exp \left\{ { - i[{q_1}({{\bf{k}}_1} \cdot {\bf{r}} + {\phi _1}) + {q_2}({{\bf{k}}_2} \cdot {\bf{r}} + {\phi _2})]} \right\}\left| {{m_1} + {q_1},{m_2} + {q_2}} \right\rangle ,
\end{array}
\end{equation}
where ${V_e}$ is the normalized volume, ${{\bf{P}}_{f}}$ is the final state momentum of the ionized electron, ${u_{{P_s}}} = \Lambda _{{P_s}}^2/{\omega _s}$ is the ponderomotive energy in units of the photon energy of the laser, where ${\Lambda _{{P_s}}}{\rm{ = }}{g_{\rm{s}}}\sqrt {{n_s}} $ is the half amplitude of the classical field under the large photon number limit ${n_s} \to \infty $ and ${g_s} \to 0$ \cite{PhysRevA.44.3169,guo1992stationary}, ${{\bf{k}}_{s}}$ denotes photon momentum of the laser field for $s=1, 2$ . The total energy of the final state is
\begin{equation}\label{eq9}
{E_{{{\bf{P}}_{f}}{m_1}{m_2}}} = \frac{{{{\bf{P}}_{f}}^2}}{2} + ({m_1} + \frac{1}{2}){\omega _1} + ({m_2} + \frac{1}{2}){\omega _2} + {u_{{P_1}}}{\omega _1} + {u_{{p_2}}}{\omega _2},
\end{equation}
where ${m_1}({m_2})$ is the photon number in the laser field. The Bessel function ${\Im _{{q_1},{q_2}}}(\zeta )$  can be written as \cite{guo1992stationary}
\begin{equation}\label{eq10}
\begin{array}{l}
\fl {\Im _{{q_1},{q_2}}}(\zeta ) = \mathop \sum \limits_{{q_3}{q_4}{q_5}{q_6}} {J_{ - {q_1} + 2{q_3} + {q_5} + {q_6}}}({\zeta _1}){e^{ - i(2{q_3} + {q_5} + {q_6}){\phi _1}}}{J_{ - {q_2} + 2{q_4} + {q_5} - {q_6}}}({\zeta _2}){e^{ - i(2{q_3} + {q_5} - {q_6}){\phi _2}}}\\
{\kern 1pt} {\kern 1pt} {\kern 1pt} {\kern 1pt} {\kern 1pt} {\kern 1pt} {\kern 1pt} {\kern 1pt} {\kern 1pt} {\kern 1pt} {\kern 1pt} {\kern 1pt} {\kern 1pt} {\kern 1pt} {\kern 1pt} {\kern 1pt} {\kern 1pt} {\kern 1pt} {\kern 1pt} {\kern 1pt} {\kern 1pt} {\kern 1pt} {\kern 1pt} {\kern 1pt} {\kern 1pt} {\kern 1pt} {\kern 1pt} {J_{ - {q_3}}}({\zeta _3}){e^{i{q_3}{\phi _3}}}{J_{ - {q_4}}}({\zeta _4}){e^{i{q_4}{\phi _4}}}{J_{ - {q_5}}}({\zeta _5}){e^{i{q_5}{\phi _5}}}{J_{ - {q_6}}}({\zeta _6}){e^{i{q_6}{\phi _6}}},
\end{array}
\end{equation}
with $\zeta  \equiv ({\zeta _1},{\zeta _2},{\zeta _3},{\zeta _4},{\zeta _5},{\zeta _6})$
\begin{equation}\label{eq11}
\fl \begin{array}{l}
{\zeta _1} = 2\sqrt {\frac{{{u_{P1}}}}{{{\omega _1}}}} {\rm{|}}\textbf{P}_{{f}} \cdot {\hat\varepsilon _1}{\rm{|}}, {\rm{     }}{\phi _1} = {\tan ^{ - 1}}\left[ {({{\rm P}_z}/{{\rm P}_x})\tan ({\xi _1}/2)} \right] + \frac{1}{2}{\Theta_1},\\
{\zeta _2} = 2\sqrt {\frac{{{u_{P2}}}}{{{\omega _2}}}} {\rm{|}}\textbf{P}_{{f}} \cdot {\hat\varepsilon _2}{\rm{|}},{\rm{    }}{\phi _2} = {\tan ^{ - 1}}\left[ {({{\rm P}_z}/{{\rm P}_x})\tan ({\xi _2}/2)} \right] + \frac{1}{2}{\Theta_2},\\
{\zeta _3} = \frac{1}{2}{u_{P_1}}\cos ({\xi _1}),{\rm{      }}{\phi _3} ={\Theta _1},\\
{\zeta _4} = \frac{1}{2}{u_{P_2}}\cos ({\xi _2}),{\rm{      }}{\phi _4} ={\Theta _2},\\
{\zeta _5} = 2\frac{{\sqrt {{u_{P_1}}{\omega _1}{u_{P_2}}{\omega _2}} }}{{{\omega _1} + {\omega _2}}}\cos \left[ {\frac{1}{2}({\xi _1} + {\xi _2})} \right],{\rm{   }}{\phi _5} = \frac{1}{2}({\Theta _1} + {\Theta _2}),\\
{\zeta _6} = 2\frac{{\sqrt {{u_{P1}}{\omega _1}{u_{P2}}{\omega _2}} }}{{\left| {{\omega _1} - {\omega _2}} \right|}}\cos \left[ {\frac{1}{2}({\xi _1} - {\xi _2})} \right],{\rm{   }}{\phi _6} = \left\{ {\begin{array}{*{20}{c}}
{\frac{1}{2}({\Theta _1} - {\Theta _2}),       {\rm{      }}{\omega _{\rm{1}}} > {\omega _2}},\\
{\frac{1}{2}({\Theta _1} - {\Theta _2}) - \pi ,{\rm{ }}{\omega _{\rm{1}}} < {\omega _2}},\\
\end{array}} \right.
\end{array}
\end{equation}
where $\Theta _1$ and $\Theta _2$ are the initial phases of the two fields. 

Now, the transition matrix element of ATI process can be written as
\begin{equation}\label{eq12}
\fl {T_{ATI}} = \left\langle {{\psi \rm_{f}}\left| V \right|{\psi \rm_i}} \right\rangle  = {V_e}^{ - 1/2} {{\rm{[(}}{{{u}}_{{P_1}}} - {q_1}){\omega _1} + {\rm{(}}{{{u}}_{{P_2}}} - {q_2}){\omega _2}]{\Im _{{q_1},{q_2}}}({\zeta _f}) {e^{ - i({q_1}{\phi _1} + {q_2}{\phi _2})}}{\Phi _{i}}({\bf{P}})}, 
\end{equation}
with ${q_1} = {l_1} - {m_1}$ being the number of the IR photons and ${q_2} = {l_2} - {m_2}$ the number of XUV photons that the electrons absorb from laser fields. ${\Phi _{i}}({\bf{P}})$ is the Fourier transform of the initial wave function ${\Phi _{i}}({\bf{r}})$. In this paper, we will mainly focus on the ATI process, because the contribution of the HATI spectrum to molecular imaging can be ignored under our computational condition.

Before presenting our numerical results, we now provide a brief discussion about the frequency-domain quantized-field description employed in this work as compared with the time-domain classical-field description. Firstly, the reason that the laser field is described by Fock states in our frequency-domain theory can be explained as follows: Since under the strong-field approximation \cite{PhysRevA.44.3169, PhysRevA.49.2117} where the photon number in the laser field can be regarded as infinity, i.e. the large photon number approximation is available  \cite{PhysRevA.44.3169, guo1992stationary}, the final state of the atom-laser system can be expressed by the quantized-field Volkov state \cite{PhysRevA.40.4997,guo1992stationary}, which is the eigenstates of the Dirac equation of an electron in a laser field under the non-relativistic approximation \cite{guo1988quantum, PhysRevA.40.4997}. This quantized-field Volkov state is an entangled state of an electron with momentum \textbf{P} and a laser field which is described by a Fock state, where the total energy of the atom-laser system is determined by the photon number in the Fock state and the electron momentum; On the other hand, by employing the Fock state of the laser field in this theory, the atom-laser system can be treated as an isolated system where its total energy keeps constant, hence we may directly derive the transition formula by using the energy conservation law for the atom-laser system. Secondly, as we mentioned in the introduction, that the S-matrix formula based on solving the time-domain Schrödinger equation with a classical field had been obtained directly by the transition equation of the frequency-domain theory for HHG processes \cite{fu2001interrelation} and HATI \cite{wang2007hati}. Additionally, the establishment of a full-quantum description has aroused rapidly increasing attention in the strong-field  \cite{shi2022explanation, PhysRevLett.130.253201, bhattacharya2023strong, gorlach2023high}. For example, A. Gorlach et al \cite{gorlach2023high}
found that the defining spectral characteristics of HHG, such as the plateau and cutoff, remain unchanged as the laser field is treated as a Fock state, which agrees with our previous results \cite{fu2001interrelation, gao2000nonperturbative}.

\section{Numerical results}
\begin{figure}[H]
  \centering	
  
  \includegraphics[width=14cm]{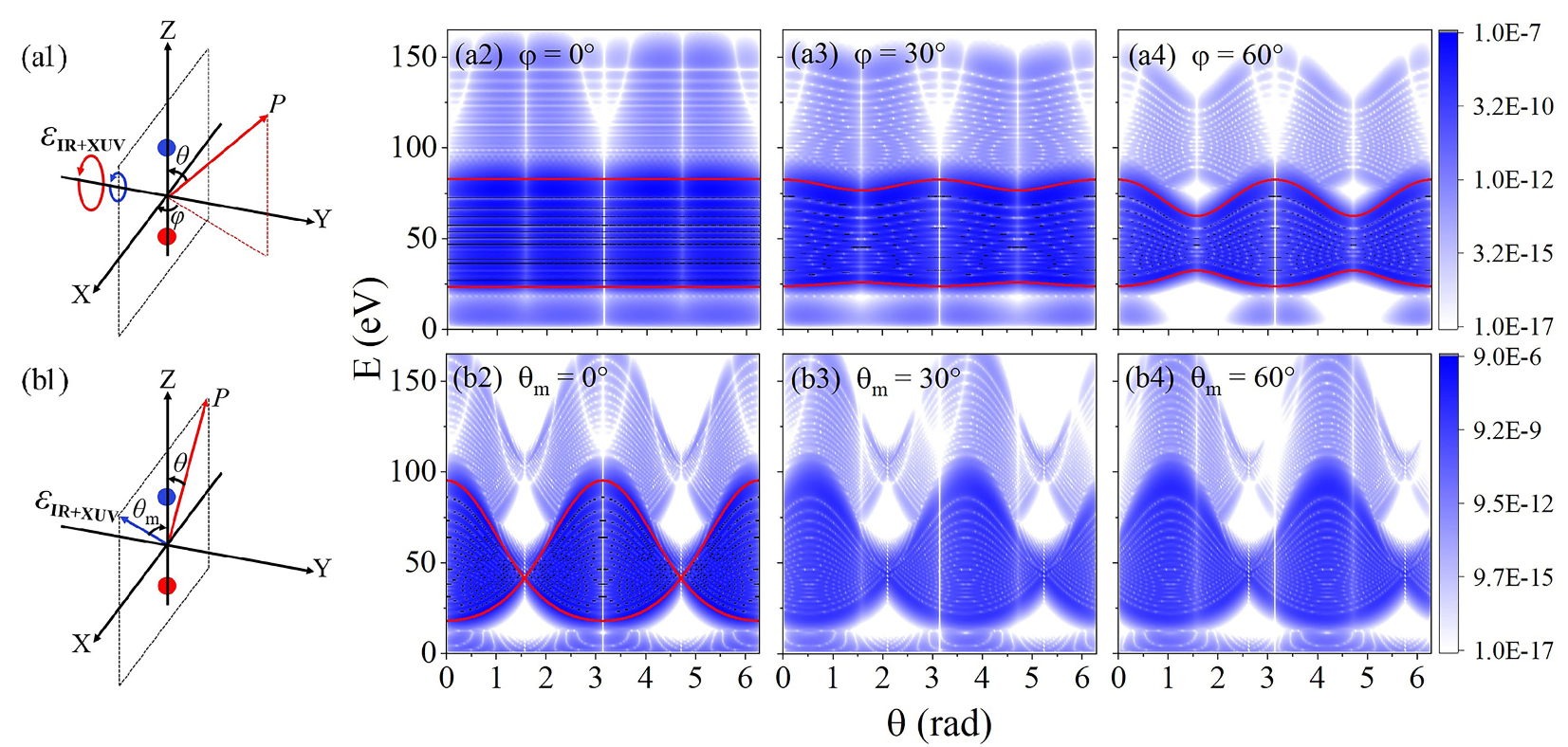}\centering
 \caption{The coordinate systems of photoelectron emission in (a1) CoRTC fields and (b1) two-color linearly polarized laser fields. In these two coordinate systems, $\theta$ is the polar angle and $\varphi $ is the azimuth angle of the photoelectron emission. In CoRCT fields, the direct angle-resolved ATI spectrum of NO with (a2) $\varphi {\rm{ = 0}}^\circ$, (a3) $\varphi {\rm{ = 30}}^\circ$ and (a4) $\varphi {\rm{ = 60}}^\circ$. In two-color linearly polarized laser fields, the direct angle-resolved ATI spectrum of NO with $\varphi {\rm{ = 0}}^\circ$ and the angle between the polarization direction of laser field and the molecular axis (b2) ${\theta _{\rm{m}}}{\rm{ = 0}}^\circ$, (b3) ${\theta _{\rm{m}}}{\rm{ = 30}}^\circ$ and (b4) ${\theta _{\rm{m}}}{\rm{ = 60}}^\circ$. In logarithmic scale.}
  \label{fig1}
\end{figure}

In this section, the ATI process of NO molecule in IR+XUV CoRTC fields is considered. We used the B3LYP/6-3111G* method implemented by the Gaussian software \cite{frisch2004gaussian} to calculate the highest occupied molecular orbital (HOMO) wave function of NO molecule in coordinate space, then we obtained the HOMO of the molecule in momentum space, according to the Fourier transform formula. The polarization planes of the CoRTC fields are set in the $xz$ plane, the frequencies of the fields are ${\omega _{\rm{1}}}{\rm{ = 1}}{\rm{.165 eV}}$, ${\omega _2} = 50{\omega _{\rm{1}}}$, and the intensities are ${I_1} = {I_2} = 3.6 \times {10^{13}}\rm W/c{m^2}$. For the sake of simplicity, the initial phases of the two fields are set at zero. Under above laser conditions, we found that the ionization rate of HATI is at least one order of magnitude smaller than that of direct ATI in the highest plateau of the spectra, so the contribution of the recollision term can be ignored. The geometry used in our calculations is illustrated in figure~\ref{fig1}, where $\theta$ is the polar angle and $\varphi $ is the azimuth angle of the ionized electron emission. The molecular axis of NO is along the $z$-axis. 

Figures~\ref{fig1}(a2)-(a4) show the angle-resolved ATI spectrum of NO molecule in CoRTC fields with different azimuth angle $\varphi $. One may find that the ATI spectrum shows a multi-plateau structure, where the widths of all the plateaus are independent of $\theta $ as $\varphi {\rm{ = 0}}^\circ$ and they oscillate with $\theta $ as $\varphi  \ne {\rm{0}}^\circ$. According to our previous work \cite{zhang2013terracelike,jin2016angle,jin2018angle,zhou2020interference}, the IR and XUV laser fields play different roles in causing the multi-plateau structure in the ATI spectrum: the XUV laser field determines the relative height of the plateaus by the number of the XUV photons that the atom absorbs, while the IR laser field determines the width of the plateaus by its intensity. Therefore, by analyzing the number of the XUV photons that the molecule absorbs which is $q_2$ expressed in Eq.~(\ref{eq12}), we find that the three plateaus shown in figures~\ref{fig1}(a2)-(a4) correspond to the XUV photon number  ${q_2} = 0,1$ and 2 absorbed by NO molecule, where the highest plateau corresponds to ${q_2} = 1$ under our present laser conditions.

We now consider the interference fringes shown in the spectrum of figures~\ref{fig1}(a2)-(a4). From the expression of the ATI transition matrix formula Eq.~(\ref{eq12}), we may classify the interference fringes in the ATI spectrum into two kinds: one kind is attributed to the interaction between the ionized electron and the laser field, i.e., the interference expressed by the Bessel function in Eq.~(\ref{eq12}), where we name these fringes as "background fringes"; while the other one is attributed to the structure of the initial molecular orbital, i.e., the fringes from the wave function of the molecule expressed in Eq.~(\ref{eq12}), where we name these fringes as "structure fringes ". Therefore, we may obtain the density distribution of the initial molecular orbital by removing the background fringes in the ATI spectrum. In the following, we first analyze the interference fringes from the Bessel function in Eq.~(\ref{eq12}).

In the CoRTC field, the polarization degree of the laser field ${\xi _1} = {\xi _2} = \pi /2$ or $3\pi /2$, hence one may find that the arguments ${\zeta _3}={\zeta _4}={\zeta _5}=0$ in the Bessel function in Eq.~(\ref{eq12}). Furthermore, under the present laser conditions in this work, the value of the argument ${\zeta _6}$ is much smaller than one, hence we may set ${J_{ - {q_6}}}({\zeta _6}){e^{i{q_6}{\phi _6}}} \approx 1.0$  and ${q_{\rm{6}}}{\rm{ = 0}}$. Therefore, the Bessel function can be reduced to
\begin{equation}\label{eq13}
{\Im _{{q_1},{q_2}}}({\zeta _f}) \approx {J_{ - {q_1}}}({\zeta _1}){J_{ - {q_2}}}({\zeta _2}){e^{i({q_1} + {q_2}){\phi _1}}}.
\end{equation}
One can find that ${J_{ - {q_1}}}({\zeta _1}){e^{i{q_1}{\phi _1}}}$ is related to the IR laser field, and ${J_{ - {q_2}}}({\zeta _2}){e^{i{q_2}{\phi _1}}}$ is related to the XUV laser field. We now discuss the characters of each plateau caused by the IR laser field, which is determined by the Bessel function ${J_{ - {q_1}}}({\zeta _1}){e^{i{q_1}{\phi _1}}}$. This Bessel function is named as phase Bessel function \cite{hu2008phased}, which is defined by ${X_n}(z) = {X_n}(x{e^{i\phi }}){\rm{ = }}{J_n}(x){e^{in\phi }}$, where the complex variable $z = x{e^{i\phi }}$ and \emph{x} is a positive number. 
The general Bessel function satisfies the following equation ${e^{ix\sin \theta }} = \sum\limits_n {{J_n}} (x){e^{in\sin \theta }}$, hence we can obtain the equation ${e^{ix\sin (\theta {\rm{ + }}\phi )}} = \sum\limits_n {{J_n}} (x){e^{in\sin \theta }}{e^{in\phi }}$ for the phase Bessel function. According to the general Bessel function integral expression, we can obtain the corresponding integral expression of the phased Bessel function:${X_n}(z) = {J_n}(x){e^{in\phi }} = \frac{1}{2}\int_{ - \pi }^\pi  {d\theta \exp \left\{ {i\left[ {x\sin (\theta  + \phi ) - n\theta } \right]} \right\}} $. Therefore, the function ${J_{ - {q_1}}}({\zeta _1}){e^{i{q_1}{\phi _1}}}$ can be expressed as 
\begin{equation}\label{eq14}
{J_{ - q_1}}({\zeta _1}){e^{i{q_1}{\phi _1}}} = \frac{1}{T}\int_{ - T/2}^{T/2} {\exp \left\{ {i\left[ {{\zeta _1}\sin ({\omega _1}t - {\phi _1}) - {q_1}{\omega _1}t} \right]} \right\}} dt,\
\end{equation}
where $T{\rm{ = }}\frac{{{\rm{2}}\pi }}{{{\omega _1}}}$. We may regard the IR laser field as a classical field, where the vector potential is along the polarization direction ${\textbf{A}_{cl}}(t) = \sqrt {{U_{P_1}}} [{\hat \varepsilon _1}{e^{ - i{\omega _1}t}} + c.c.]$. Therefore, the classical action of an electron in the IR laser field is
\begin{equation}\label{eq15}
\fl {S_{c1}}({\textbf{P} _{f}},t) = \frac{1}{{2{m_e}}}{\int_0^t {\left[ {{\textbf{P}_{f}} + e{\textbf{A}_{cl}}(t')} \right]} ^2} dt'= (\frac{{{\textbf{P} _{f}^2}}}{{2{m_e}}} + {U_{P_1}})t + {\zeta _{_1}}\sin ({\omega _1}t - {\phi _1}).
\end{equation}
Using the above classical action formula, the Bessel function can be rewritten as
\begin{equation}\label{eq16}
{J_{ - {q_1}}}({\zeta _1}){e^{i{q_1}{\phi _1}}} = \frac{1}{T}\int_{ - T/2}^{T/2} {{e^{if(t)}}}dt,
\end{equation}
where $f(t) = {S_{c1}}({\textbf{P}_{f}},t) - ({q_2}{\omega _2} - {I_P})t$ with $\frac{{{\textbf{P}_{f}^2}}}{{2{m_e}}} + {U_{P_1}}={q_1}{\omega _1}+{q_2}{\omega _2} - {I_P}$ for energy conversation during the ionization process. Using the saddle-point approximation\cite{wang2007hati}, Eq.~(\ref{eq16}) becomes
\begin{equation}\label{eq17}
{J_{ - {q_1}}}({\zeta _1}){e^{i{q_1}{\phi _1}}} = \frac{{2{\omega _1}}}{{\sqrt {\pi f''({t_0})} }}\cos \left[ {f({t_0}) - \frac{\pi }{4}} \right].
\end{equation}
Based on Eq.~(\ref{eq17}), the interference is attributed to the term $\cos \left[ {f({t_0}) - \frac{\pi }{4}} \right]$, where the minimum value in the spectrum occurs as $\cos \left[ {f({t_0}) - \frac{\pi }{4}} \right] = 0$. For example, the interference fringes for the highest plateau, where $q_2=1$, in the spectrum in figures~\ref{fig1}(a2)-(a4), are predicted by black dot dots. It can be seen that these fringes are in good agreement with the positions of destructive interference fringes obtained by the numerical calculation. Furthermore, the saddle-point ${t_0}$ satisfies ${f^\prime }(t)\left| {_{t = {t_0}}} \right. = 0$, leading to the following equation:
\begin{equation}\label{eq18}
\frac{{{{\left[ {{\textbf{P}_{f}} + {\textbf{A}_{cl}}({t_0})} \right]}^2}}}{2} = {q_2}{\omega _2} - {I_P}.
\end{equation}
This equation expresses the energy conservation when the electron is ionized from the bound state into the continuum at time ${t_0}$ in the IR laser field with absorbing ${q_2}$ XUV photons. In addition, Eq.~(\ref{eq18}) may predict the beginning and cutoff positions of the ATI spectrum for the plateau in the spectrum, where the maximum energy value is
\begin{equation}\label{eq19}
\fl {E_{\max }} \! = \! \frac{{{{\left( \!\! {\sqrt {2({q_2}{\omega _2} \! - \! {I_P} \! - \! {U_{P1}}) \! + \! 2{U_{P1}}({{\sin }^2}\theta {{\cos }^2}\varphi \! + \! {{\cos }^2}\theta )} \! + \! \sqrt {2{U_{P1}}({{\sin }^2}\theta {{\cos }^2}\varphi \! + \! {{\cos }^2}\theta )} } \right )}^2}}}{2},
\end{equation}
and the minimum energy value is
\begin{equation}\label{eq20}
\fl {E_{\min }} \! = \! \frac{{{{\left(\!\! {\sqrt {2({q_2}{\omega _2} \! - \! {I_P} \! - \! {U_{P1}}) \! + \! 2{U_{P1}}({{\sin }^2}\theta {{\cos }^2}\varphi \! + \! {{\cos }^2}\theta )} \! - \! \sqrt {2{U_{P1}}({{\sin }^2}\theta {{\cos }^2}\varphi \! + \! {{\cos }^2}\theta )} } \right)}^2}}}{2},
\end{equation}
where for the case that $q_2\omega_2<I_P$ the minimum ( i.e., beginning) position is at zero \cite{zhang2013terracelike}. Specially, when $\varphi {\rm{ = 0}}^\circ$, the equations reduce into 
${E_{\max }} = {\left( {\sqrt {2({q_2}{\omega _2} - {I_P})} + \sqrt {2{U_{P1}}} } \right)^2}/2$ and ${E_{\min }} = {\left( {\sqrt {2({q_2}{\omega _2} - {I_P})} - \sqrt {2{U_{P1}}} } \right)^2}/2$ , which are independent of the angle $\theta$. The red horizontal lines in figure~\ref{fig1}(a2) predict the beginning and cutoff positions of the plateau by the above equations for ${q_2}=1$, which shows that the numerical results agree very well with the prediction of Eqs.~(\ref{eq19}) and (\ref{eq20}).

For the sake of comparison, figures~\ref{fig1}(b2)-(b4) show the angle-resolved ATI spectrum in IR+XUV two-color linearly polarized laser fields, where the polarization direction of the laser field is shown in figure~\ref{fig1}(b2) with the angle $\theta_m$, which is the angle between the polarization direction of laser field and the
molecular axis, and the intensities and frequencies of the two-color line polarized field are the same as that of the CoRTC field. Comparing with figures~\ref{fig1}(a2)-(a4), figures~\ref{fig1}(b2)-(b4) show that the width of the plateau in the spectrum oscillates with the angle $\theta$ and the whole pattern of the background fringes in the spectrum shifts with the direction of the two-color laser polarization. Especially, the background fringes show a more complex structure, where these fringes not only depend on the direction of the electron emission but also the direction of the laser polarization. The prediction of the background fringes is also presented in figure~\ref{fig1}(b2) for $q_2=1$ by analyzing the corresponding Bessel function, and the beginning and cutoff positions of this plateau were shown by red lines using formula ${E_{\min }} = {\left(\sqrt {2({q_2}{\omega_2} - {I_P})+4U_{P_1}\sin^{2}(\theta)}- 2\sqrt {U_{P_1}\cos^2(\theta)}\right)^2}/2$ and ${E_{\max }} = {\left(\sqrt {2({q_2}{\omega_2} - {I_P})+4U_{P_1}\sin^{2}(\theta)}+ 2\sqrt {U_{P_1}\cos^2(\theta)}\right)^2}/2$ respectively. These comparison results indicate that the CoRTC laser field is a more convenient tool than the linearly polarized laser field for us to image molecular structures by ionization spectrum. This is because that the ionization spectrum depends on the polarization direction $\theta_m$, hence we need to scan the polarization direction to provide a complete imaging of the molecular orbitals in linearly polarized laser fields.

\begin{figure}[H]
  \centering
  \vspace{-0.3cm}   
\setlength{\abovecaptionskip}{-0.3cm} 
\setlength{\abovecaptionskip}{0cm} 
  \includegraphics[width=12cm]{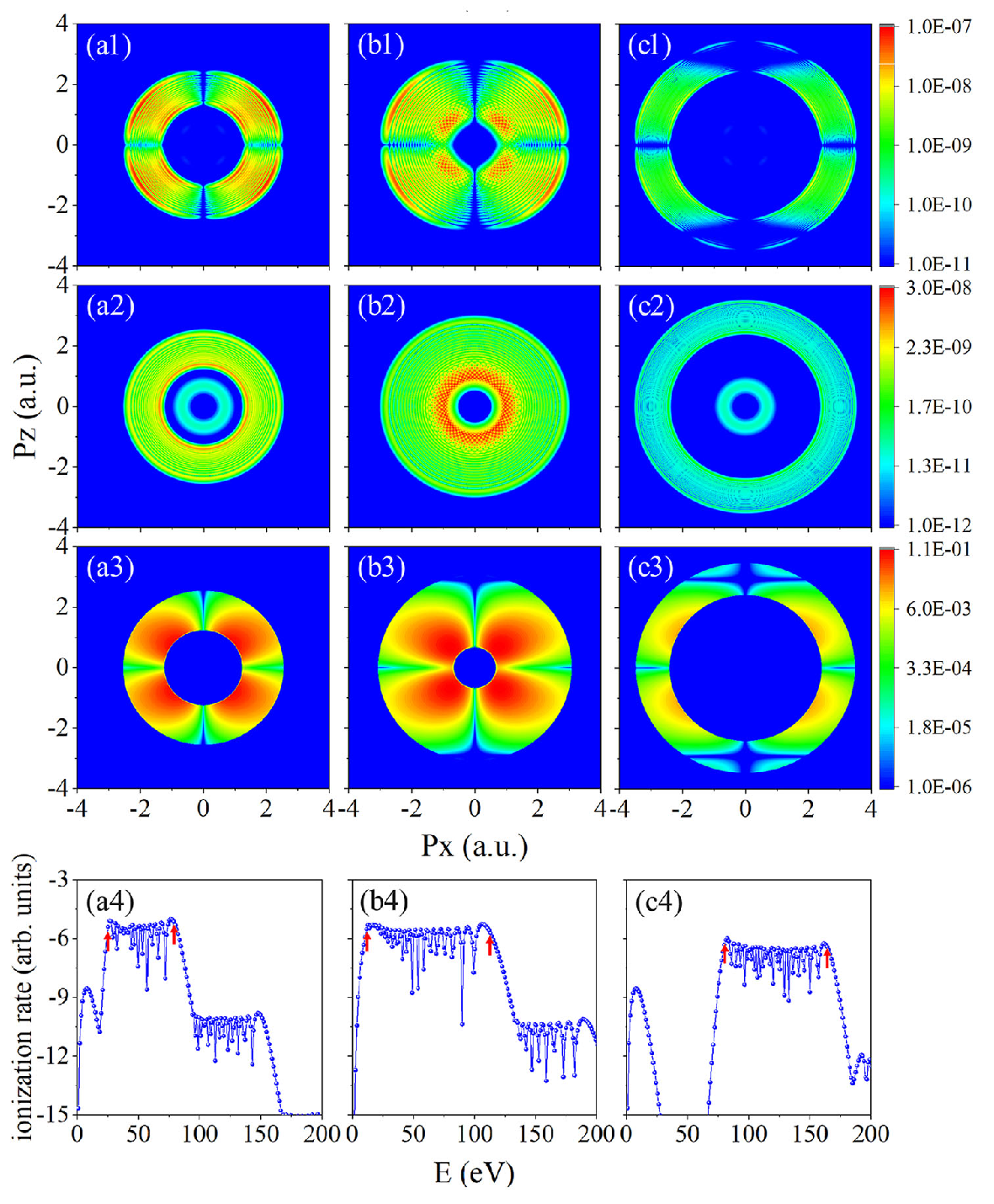}\centering
   \caption{ATI momentum spectra of NO (first line: (a1), (b1) and (c1)) and the atom with the same ionization energy as NO (second line: (a2), (b2) and (c2)). The density distribution of HOMO of NO (third line: (a3), (b3), and (c3)) in momentum space is obtained by the corresponding ATI momentum spectrum. The direct ATI spectra of NO in CoRTC fields (Fourth line: (a4), (b4) and (c4)). The frequencies and intensities of the CoRTC fields are (first column: (a1)-(a4)) ${\omega _{\rm{1}}}{\rm{ = 1}}{\rm{.165 eV}}$, ${\omega _2} = 50{\omega _{\rm{1}}}$, and ${I_1} = {I_2} = 3.6 \times {10^{13}} \rm W/c{m^2}$, (second column: (b1)-(b4)) ${\omega _{\rm{1}}}{\rm{ = 1}}{\rm{.165 eV}}$, ${\omega _2} = 50{\omega _{\rm{1}}}$  and ${I_1} = 1.2 \times {10^{14}} \rm W/c{m^2}$, ${I_2} = 3.6 \times {10^{13}} \rm W/c{m^2}$, (third column: (c1)-(a4)) ${\omega _{\rm{1}}}{\rm{ = 1}}{\rm{.165 eV}}$, ${\omega _2} = 110{\omega _{\rm{1}}}$, and ${I_1} = {I_2} = 3.6 \times {10^{13}}\rm W/c{m^2}$. In logarithmic scale.}
    \label{fig2}
\end{figure}

We now consider how to image the density distribution of the molecule wave function by using the ionization spectrum in CoRTC laser fields. As we mentioned above, the interference fringes in the ATI spectrum of a molecule include two kinds: the background fringes and the structure fringes. In order to image the molecular orbital by the direct ATI energy spectrum, the background interference fringes in the energy spectrum should be eliminated. According to Eq.~(\ref{eq12}), we can obtain ${T_{ATI}} \propto {\Im _{{q_1},{q_2}}}({\zeta _f}) \times {\Phi _i}({\bf{P}})$, and find that background fringes that come from the Bessel function ${\Im _{{q_1},{q_2}}}({\zeta _f})$ only depend on the laser conditions and the ionization energy of the molecule, hence under the same laser conditions, the values of the Bessel function for an atom are same as that for a molecule with the same ionization energy. Therefore, similar to the method of extracting the recovery dipole moment from the high-order harmonic spectrum  \cite{itatani2004tomographic,le2009quantitative}, we obtain the molecular wave function by the formula: ${\left| {\Phi _i^{mol}({\bf{P}})} \right|^{\rm{2}}}{\rm{ = }}\frac{{{{\left| {T_{ATI}^{mol}} \right|}^{\rm{2}}}}}{{{{\left| {T_{ATI}^{ref}} \right|}^{\rm{2}}}}}{\left| {\Phi _i^{ref}({\bf{P}})} \right|^{\rm{2}}}$, where $T_{ATI}^{mol}$ and $T_{ATI}^{ref}$ are transition matrix elements of NO molecule and the atom with the same ionization energy as NO molecule respectively, $\Phi _i^{mol}({\bf{P}})$ and $\Phi _i^{ref}({\bf{P}})$ are wave functions in the momentum space of NO molecule and the atom with the same ionization energy as NO molecule respectively. The practical steps are as follows: At first, we calculated the ATI spectrum of hydrogen-like atoms with the same ionization energy as NO molecule under the same laser conditions as shown in figure~\ref{fig2}(a2), where it can be seen that the interference fringes in figure~\ref{fig2}(a2) are consistent with the background fringes in the spectrum of NO molecule shown in figure~\ref{fig2}(a1). Then we divide the data of the ATI momentum spectrum of NO molecule by that of the hydrogen-like atom and multiply the density distribution of the atomic initial state in momentum space. At last, the structure fringe from the ATI spectrum caused by the HOMO of NO molecule is obtained, as shown in figure~\ref{fig2}(a3), which agrees with the density distribution of the wave function of NO molecule.

Furthermore, as shown in figure~\ref{fig2}(a4), which is the total ATI spectrum obtained by integrating the polar angles $\theta $ of the photoelectron emission, we may find that the ionization rate of the highest plateau ($q_2=1$) is larger than that of the other plateaus more than three orders of magnitude. This indicates that only the highest plateau plays a dominant role in imaging the molecular structure, as shown in figure~\ref{fig2}(a3). Therefore, in order to image the density distribution of the initial molecular state effectively, we may change the position and width of the highest plateau in the ATI spectrum by applying different laser conditions, hence the range of the highest plateau of the spectrum may cover the range of the density distribution of the molecular orbital in momentum space. To illustrate this proposal, figures~\ref{fig2}((a4), (b4), and (c4)) show the ATI spectrum in the CoRCT fields for ${\omega _{\rm{1}}}{\rm{ = 1}}{\rm{.165 eV}}$ with (a4) ${\omega _2} = 50{\omega _{\rm{1}}}$ and laser intensities of ${I_1} = {I_2} = 3.6 \times {10^{13}}\rm W/c{m^2}$, (b4) ${\omega _2} = 50{\omega _{\rm{1}}}$ and laser intensities of ${I_1} = 1.2 \times {10^{14}}\rm W/c{m^2}$ and ${I_2} = 3.6 \times {10^{13}} \rm W/c{m^2}$, and (c4) ${\omega _2} = 110{\omega _{\rm{1}}}$ and laser intensities of ${I_1} = {I_2} = 3.6 \times {10^{13}}\rm W/c{m^2}$. As shown in figure~\ref{fig2}(a4), one may find that the energy range of the highest plateau ($q_2=1$) is 25-80eV, and the positions of the maximum and minimum energy of this plateau are marked with arrows, hence the corresponding range of the momentum distribution of the ionized electron is between 1.4-2.4 a.u., where the corresponding density distribution of the molecular state is shown in figure~\ref{fig2}(a3). However, the width of the highest plateau in the ATI spectrum increases with range 12-115eV as the IR laser intensity increases to $1.2 \times {10^{14}} \rm W/c{m^2}$ as shown in figure~\ref{fig2}(b4), hence the corresponding range of the momentum distribution enlarges, and as a result, a more complete electron density distribution of the initial molecular wave function in momentum space can be obtained, as shown in figure~\ref{fig2}(b3).

\begin{figure}[H]
  \centering
  \vspace{-0.3cm}   
\setlength{\abovecaptionskip}{-0.3cm} 
\setlength{\abovecaptionskip}{0cm} 
 \includegraphics[width=12cm]{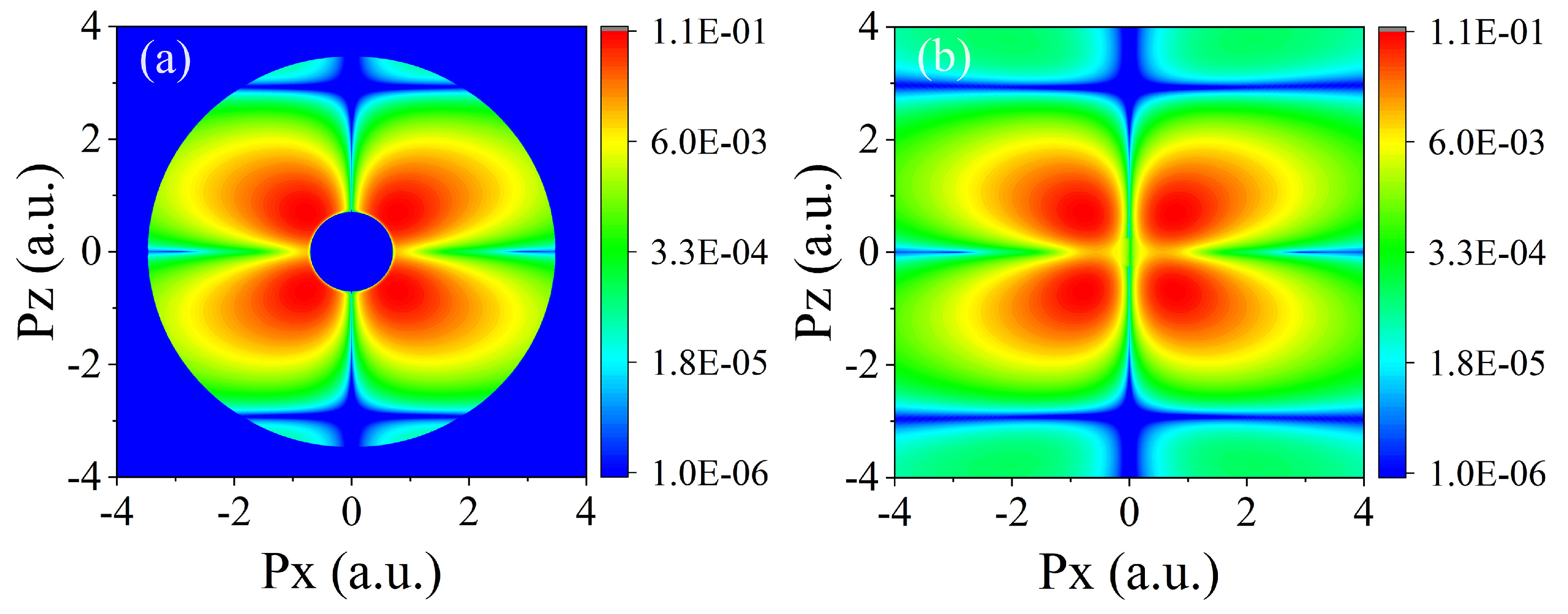}\centering
   \caption{The density distribution of HOMO of NO in momentum space (a) imaged by the ATI momentum spectrum and (b) simulated by quantum chemistry software.}
    \label{fig3}
\end{figure}

In addition,  as shown in Ref \cite{zhou2020interference}, we know that the interference fringes in the density distribution in the high momentum region of the electron may image the molecular structure. In order to obtain the molecular structure by this method, we present the ATI spectrum of the electron with a range of 80-165 eV as shown in figure~\ref{fig2}(c1), where the corresponding density distribution of the molecule orbital illustrates the fringes coming from the molecular structure shown in figure~\ref{fig2}(c3). Based on these fringes the bond length of the NO molecule can be obtained as follows: Since the HOMO orbital in the momentum space of NO molecule can be approximately expressed as $\Phi ({\bf{P}}) \propto {P_{\rm{x}}}\sin ({R_{NO}}{P_{\rm{z}}})$, the destructive interference fringes occur as $\sin ({R_{NO}}{P_{\rm{z}}}) = 0$, hence we can obtain the molecular bond length by ${R_{NO}} = \frac{{ \pm n\pi }}{{P_{\rm{z}}}}$ with the fringes at $P_{\rm{z}}$. It can be seen from figure~\ref{fig2}(c3) that the value of ${P_{\rm{z}}}$ at the interference fringe is $ \pm {\rm{2}}{\rm{.96 a}}{\rm{.u}}{\rm{.}}$, hence by using the above formula, we can obtain that the bond length of NO molecule is 1.12{\AA}, which is very close to the value provided in Ref. \cite{final}. Finally, combining figures~\ref{fig2} ((a3), (b3), and (c3)), we may obtain the whole NO molecular orbital in momentum space, as shown in figure~\ref{fig3} (a). In figure~\ref{fig3} (b), we give the electron density distribution of the HOMO of NO molecule obtained by quantum chemistry software.

Comparing figures~\ref{fig3} (a) and (b), one may find that the electron density distribution of the HOMO of NO molecules obtained by the ATI momentum spectrum is consistent with that simulated by quantum chemistry software. In addition, we choose the appropriate energy region of the ATI spectrum to image the HOMO of the molecules by using suitable two-color laser conditions in order to avoid the error in the low energy region of the ATI spectrum caused by the strong-field approximation in the frequency-domain theory.

\begin{figure}[H]
\centering
  \includegraphics[width=12cm]{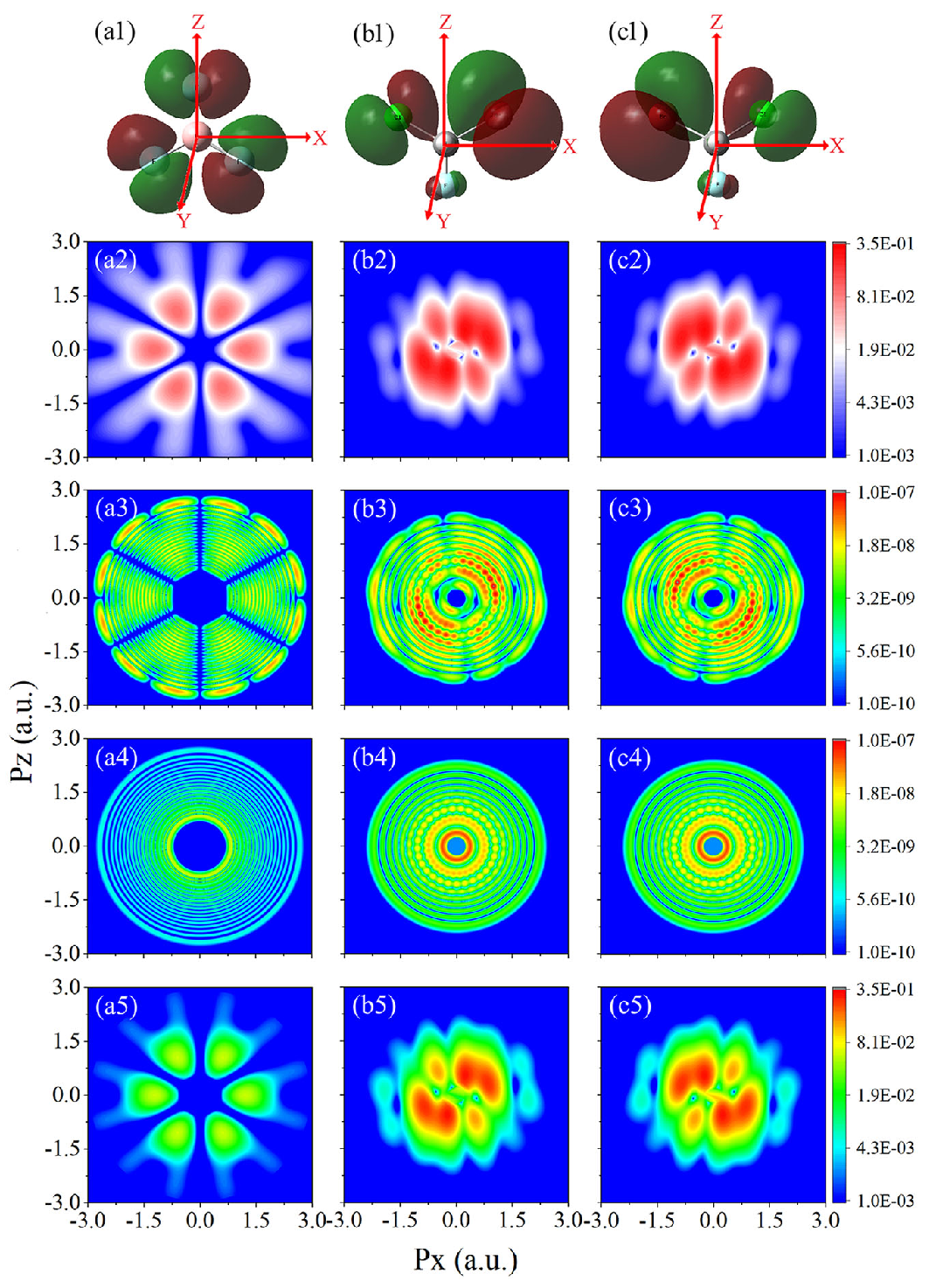}
  \caption{The HOMO of (a1) $ \rm BF_3 $, (b1) R-CFClBrH and (c1) S-CFClBrH, the HOMO electron density distribution of (a2) $\rm BF_3$, (b2) R-CFClBrH and (c2) S-CFClBrH in momentum space, the ATI momentum spectrum of (a3) $ \rm BF_3$, (b3)R-CFClBrH and (c3) S-CFClBrH, the ATI momentum spectrum of the atom with the same ionization energy as (a4) $ \rm BF_3$, (b4) R-CFClBrH and (c4) S-CFClBrH, the density distribution of HOMO of (a5) $ \rm BF_3$, (b5) R-CFClBrH and (c5) S-CFClBrH in momentum space obtained by the ATI momentum spectrum of the $\rm BF_3$, R-CFClBrH and S-CFClBrH, separately. The frequencies and intensities of the CoRTC fields are ((a3), (a4)) ${\omega _{\rm{1}}}{\rm{ = 1}}{\rm{.165 eV}}$, ${\omega _2} = 50{\omega _{\rm{1}}}$, ${I_1} = 1.2 \times {10^{14}}\rm W/c{m^2}$ and ${I_2} = 3.6 \times {10^{13}} \rm W/c{m^2}$, ((b3), (b4)) and ((c3), (a4))$ {\omega _{\rm{1}}}{\rm{ = 1}}{\rm{.165 eV}}$, ${\omega _2} = 30{\omega _{\rm{1}}}$, ${I_1} = 1.2 \times {10^{14}} \rm W/c{m^2}$ and ${I_2} = 3.6 \times {10^{13}}\rm W/c{m^2}$. In logarithmic scale. }
    \label{fig4}
\end{figure}

In general, we may control the molecular ATI spectrum in different energy ranges by adjusting the CoRTC laser field conditions for different molecules and then image the momentum distribution of the molecular wave function in a proper momentum region. In the following, we present the imaging of complex molecules using this method.  Figure~\ref{fig4} (a1) shows the HOMO and (a2) the density distribution of this orbital in momentum space of $\rm BF_3$ molecule, and figures\ref{fig4}(a3) and (a4) show the angle-resolved ATI spectrum of $\rm BF_3$ (a3) and the spectrum (a4) of the corresponding hydrogen-like atom with the same $I_P$ of $ \rm BF_3$, while figure~\ref{fig4}(a5) shows the imaging of the initial HOMO of $ \rm BF_3$ molecule by using the spectrum of figures~\ref{fig4}(a3) and (a4). In order to obtain the whole density distribution of the initial HOMO of $\rm BF_3$, here we choose the laser field with the frequencies being ${\omega _{\rm{1}}}{\rm{ = 1}}{\rm{.165 eV}}$ and ${\omega _2} = 50{\omega _{\rm{1}}}$, and the intensities ${I_1} = 1.2 \times {10^{14}} \rm W/c{m^2}$ and ${I_2} = 3.6 \times {10^{13}}\rm W/c{m^2}$. Comparing figures\ref{fig4} (a2) and (a5), one may find that the density distribution of the $ \rm BF_3$ initial state can be imaged by the angle-resolved ATI spectrum in the CoRTC laser fields.

At last, we imaging two chiral molecules by using their ATI spectrum in the CoRTC laser fields. As we know that chirality is a ubiquitous naturally occurring phenomenon that plays a major role in Physics, Chemistry, and Biology. Chiral molecules appear in pairs of left- and right-handed enantiomers, and how to distinguish them is still a vital and hot topic.  Here we show that the chirality of molecules can be resolved by using the angle-resolved ATI spectrum in IR+XUV CoRTC laser fields. Figures~\ref{fig4}(b1) and (c1) show the HOMO of R and S-CFClBrH molecules, respectively. The molecular axis for both molecules is along the $z$-axis. By using the CoRTC laser fields with ${\omega _{\rm{1}}}{\rm{ = 1}}{\rm{.165 eV}}$, ${\omega _2} = 30{\omega _{\rm{1}}}$, ${I_1} = 1.2 \times {10^{14} } \rm W/c{m^2}$ and ${I_2} = 3.6 \times {10^{13}} \rm W/c{m^2}$, we present the angle-resolved ATI spectra for R- (b3) and S- (c3) CFClBrH molecules, and the corresponding ATI spectrum of the hydrogen-like atom with the same ionization energy of R- (b4) and S- (c4) CFClBrH molecules. The imaging of the initial state density distribution is shown in figure~\ref{fig4}(b5) for R- and (c5) for S- CFClBrH molecules, where one may find that different chirality from oriented chiral molecules can be easily distinguished by this method. From all these examples, we may find that, for any unknown molecule, as long as the ionization energy of the molecule is measured \cite{turner1966ionization}, the angle-resolved ATI spectrum can be obtained by scanning the CoRTC laser conditions, and then the density distribution of the molecular HOMO can be imaged. Additionally, this proposal can also be realized by time domain method, although it may take a long calculation time for such IR+XUV laser fields case.

\section{Conclusion}
Based on the frequency-domain theory, we studied the above threshold ionization of the orientated molecules in IR+XUV co-rotating circular laser fields. It is found that the co-rotating circular laser fields are a more convenient tool to image molecular orbital by the interference fringes of angle-resolved ATI spectrum than the two-color linearly polarized laser fields. Moreover, because IR and XUV laser fields play different roles in the ionization process, we can obtain ATI momentum spectra of molecules in different momentum ranges by changing laser field conditions. Therefore, the molecular HOMO in the appropriate momentum range can be imaged. This work shed light on the study of imaging a complex molecular structure by photoelectron spectra in co-rotating circular laser fields.

\section*{Data Availability Statement}
All data that support the findings of this study are included
within the article (and any supplementary files).

\section*{Acknowledgments}
We thank all the members of SFAMP club for helpful discussions. This work was supported by the National Natural Science Foundation of China under Grant Nos. 92250303, 12074418, 12204526, 12104285,
11334009, 11425414, and the Guangdong Basic and Applied Basic Research Foundation(No. 2022A1515011742).
\section*{References}
\bibliography{ref}
\bibliographystyle{unsrt}
\end{document}